# Deep-Subwavelength Plasmon Polariton Atomic Cavity Detector for Frequency- and Polarization-Sensitive Terahertz Detection and Imaging

*Shaojing Liu[†], Hongjia Zhu[†], Ximiao Wang, Runli Li, Hai Ou, Shangdong Li, Yanlin Ke, Runze Zhan, Ningsheng Xu[*], Huanjun Chen[*], and Shaozhi Deng[*]*

S. Liu, H. Zhu, X. Wang, R. Li, H. Ou, S. Li, Y. Ke, R. Zhan, N. Xu, H. Chen, S. Deng
State Key Laboratory of Optoelectronic Materials and Technologies Guangdong Province
Key Laboratory of Display Material and Technology
School of Electronics and Information Technology
Sun Yat-sen University
Guangzhou 510275, China
E-mail: chenhj8@mail.sysu.edu.cn; stsxns@mail.sysu.edu.cn; stsdsz@mail.sysu.edu.cn

Room-temperature, miniaturized, polarization-resolved terahertz (THz) detection of high speed is vital for high-resolution imaging in radar, remote sensing, and semiconductor inspection, and is essential for large-scale THz focal plane arrays. However, miniaturization below deep-subwavelength scales (< $\lambda_0/50$) remain challenging due to weak light–matter interaction, which degrades responsivity and polarization sensitivity. Here, we present a graphene plasmon polariton atomic cavity (PPAC) monolithic detector that overcomes this limitation by maintaining and even enhancing performance at a deep-subwavelength channel length of just 2 μm ($\lambda_0/60$). The device integrates graphene rectangle PPAC arrays with dissimilar metal contacts, where graphene functions as both absorber and conductor, simplifying the architecture. Exploiting plasmon polariton resonances and the photothermoelectric (PTE) effect, the detector achieves polarization-sensitive, frequency-selective, and fast THz detection spanning 0.53 to 4.24 THz with a polarization ratio of 93, featuring a responsivity ($R_V$) of 1007 V/W, a noise-equivalent power (NEP) of 16 pW/Hz$^{0.5}$, a specific detectivity ($D^*$) of $2.9 \times 10^7$ Jones, and a response time of 230 ps. We further demonstrate monolithic integration for polarization imaging and non-destructive semiconductor chip inspection, advancing room-temperature, compact, and polarization-sensitive THz technologies.

# 1. Introduction

THz waves (0.1–10 THz), occupying the electromagnetic spectrum between infrared and microwave radiation, offer rich spectral information that enables advanced material characterization, anisotropic imaging, and high-contrast object identification.[1–3] Room-temperature polarization-resolved THz detection has emerged as a critical technology for next-generation applications, including high-resolution radar imaging, secure wireless communication, and non-invasive semiconductor inspection.[4–6] The ability to capture the polarization state of THz radiation is thus pivotal for both fundamental research and the development of applied THz systems. Moreover, the rapid progress in large-scale THz focal plane arrays has driven the demand for miniaturized, high-density integrated detectors, underscoring the need for compact nano-/micro-detectors with polarization-resolving capabilities.

However, a substantial performance gap remains between the demands of the above advanced THz detection technology and the capabilities of existing mature detectors. Thermal detectors—such as Golay cells, thermopiles, and bolometers—operate at room temperature by sensing THz-induced heating, but they typically offer slow response times on the order of microseconds to milliseconds.[7–9] Electronic detectors, including high electron mobility transistors (HEMTs) and Schottky barrier diodes (SBDs), provide fast (nanosecond-scale) and sensitive detection (21 pW/Hz$^{0.5}$ and 110 pW/Hz$^{0.5}$ for HEMT and SBD, respectively), but their operational bandwidth is narrow (typically ~0.1 THz), and their sensitivity rapidly declines above 1 THz.[10,11] Photonic detectors based on inter- or intraband transitions in narrow bandgap semiconductors offer high speed (~μs) and sensitivity (~1 pW/Hz$^{0.5}$), yet they generally require cryogenic cooling for operation.[12] Critically, conventional polarization-sensitive THz detection often depends on bulky optical components such as rotating polarizers or wave plates, leading to long acquisition times, alignment complexity, and incompatible with on-chip integration. To address these limitations, flat optics based on metamaterials and metasurfaces have been proposed for integration with THz detectors.[13,14] However, these approaches face challenges, including difficult alignment during heterogeneous integration and reduced light throughput to the photosensitive region. A monolithic strategy—free of additional optical components—that not only enables polarization selection but also enhances light absorption in the detection material is highly desirable.

Van der Waals (vdW) two-dimensional (2D) materials, represented by graphene, have garnered significant attention for advancing photodetection technologies, owing to their tunable bandgaps, gapless electronic structures, and ultrahigh carrier mobility at room temperature.[15,16]

Consequently, room-temperature THz detectors have been developed using graphene,[17–19] semimetals,[20–22] and topological insulators.[23–25] Despite notable progress, a longstanding challenge remains: the responsivity of THz detectors degrades substantially as the device footprint shrinks, primarily due to the weak interaction between long-wavelength THz waves and atomically thin active layers. Furthermore, most 2D materials exhibit weak intrinsic anisotropic THz absorption, necessitating the integration of additional polarization-selective components—such as antennas,[26,27] anisotropic absorbers,[28] or metasurfaces[5,29]—which increases device complexity and lateral size. These constraints pose major barriers to lateral miniaturization. Achieving high sensitivity alongside strong polarization selectivity within a compact footprint remains a significant unsolved challenge in THz detector design.

Recently, we introduced ***plasmon polariton atomic cavities*** (**PPACs**) by patterning monolayer graphene into periodic nanostructures that support strong surface plasmon polariton resonances (SPPRs). Building on this concept, we demonstrate monolithic THz detectors that achieve high responsivity and pronounced polarization sensitivity without requiring additional optical components.[4] In this study, we present a compact, high-performance THz detector featuring an asymmetric graphene channel composed of graphene nano-rectangles—functioning as PPACs—and unpatterned graphene, connected by two dissimilar metal electrodes. The choice of graphene rectangle PPACs is particularly advantageous for device miniaturization. Unlike circular disks as we previously reported, whose SPPR frequency is solely determined by their diameter—thus fixing the disk size once the target detection frequency is set—rectangle PPACs offer tunability through their aspect ratio (length-to-width ratio). This geometric flexibility allows for simultaneous scaling down of both dimensions while maintaining the desired resonance frequency, thereby enabling the shortening of the device channel. As a result, detector miniaturization down to deep sub-wavelength scales becomes feasible. The graphene rectangle PPACs therefore provide strong, frequency- and polarization-selective plasmonic resonances, while the asymmetric electrodes with different work functions promote directional hot-carrier transport via the PTE effect. Crucially, when the channel length is scaled below 10 μm, the PPACs and metal contacts synergistically enhance photovoltage generation. The metal electrodes concentrate the THz field into localized hotspots overlapping the active channel, thereby boosting THz wave absorption and photocurrent. As a result, the THz photovoltage remains nearly constant even as the channel length decreases. This enables a deep-subwavelength detector with a lateral footprint of only 2 μm × 16 μm (~$\lambda_0$/60) and atomic thickness. The device delivers broadband (0.53–4.24 THz), polarization-sensitive detection with a polarization ratio (*PR*) of 93, a $R_V$ of 1007 V/W (98 V/W), and an NEP of 16 pW/Hz$^{0.5}$ (165 pW/Hz$^{0.5}$), referenced to the absorbed (incident) THz power. Using this miniaturized detector, we further demonstrate polarization imaging and non-destructive inspection of the internal structures of a semiconductor chip, showcasing its potential for monolithic integrated, high-density THz detection and imaging systems.

## 2. Design, fabrication, and characterization of graphene rectangle PPAC

The confinement and absorption of the electromagnetic field in a graphene PPACs are significantly enhanced when the incident THz wave frequency matches the resonance frequency of SPPRs in the array. The absorption intensity, resonance frequency, and polarization characteristics of the plasmonic response can be tuned by varying the length ($L_r$), width ($W_r$), and the separations between adjacent PPACs ($d_w$ and $d_l$).[4] To systematically investigate the influence of these geometrical parameters on SPPRs resonance behavior, finite-difference time-domain (FDTD) simulations were performed. In the simulations, THz waves were incident perpendicularly onto the array, with the polarization angle ($\theta$) set to 0°—that is, $x$-polarized—so that the THz electric field aligned with the long axis of the PPACs (**Figure 1**a).

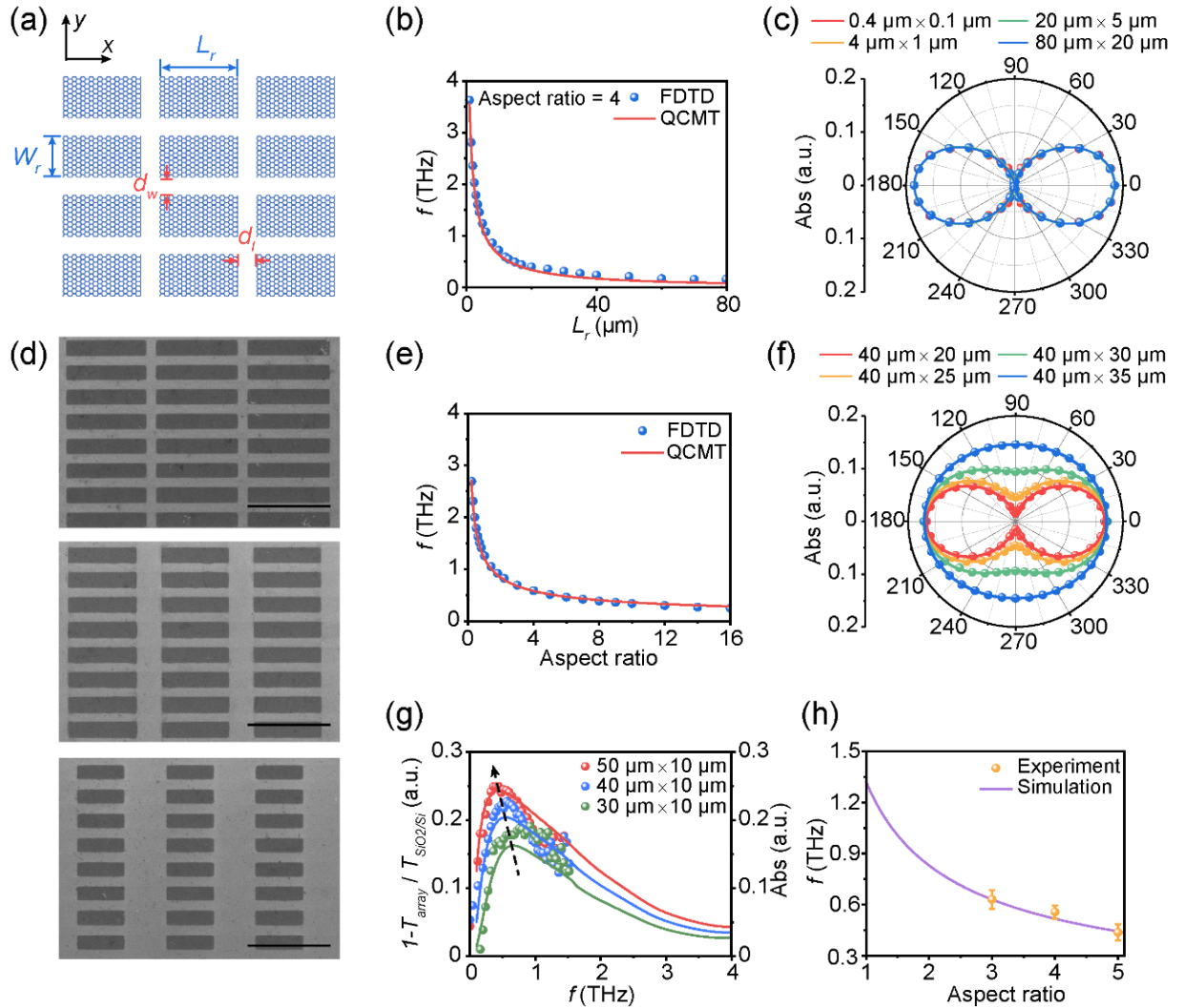

**Figure 1.** Graphene rectangle PPACs and their plasmonic resonance characteristics. (a) Schematic illustration of the graphene rectangle PPAC array. $L_r$ and $W_r$ represent the length and width of the PPAC, respectively. The inter-resonator spacing along the *x*- and *y*-axes is denoted as $d_w$ and $d_l$, respectively. (b) Simulated resonance frequency, *f*, of the PPACs as a function of $L_r$, with a fixed aspect ratio of 4 ($L_r = 4W_r$) and $L_r$ ranging from 1 μm to 80 μm. (c) Simulated polarization polar plot of the resonance for PPACs with a fixed aspect ratio of 4. The polarization angle *θ* is defined as the angle between the THz polarization direction and long axis of the PPAC. (d) SEM images of rectangle PPAC arrays with dimensions of 50 μm × 10 μm (top panel), 40 μm × 10 μm (middle panel), and 30 μm × 10 μm (bottom panel). Scale bars: 50 μm. (e) Simulated *f* of the PPACs as a function of aspect ratio varying from 0.22 to 16, with a fixed $W_r = 5$ μm. (f) Simulated polarization polar plot of the resonance for PPACs with varying aspect ratios from 1.14 to 2. (g) Experimental (dots) and simulated (lines) THz absorption spectra of the PPAC arrays in (d), showing the resonance frequency shift with changing aspect ratio. (h) Comparison of experimental (dots) and simulated (line) resonance frequencies as a function of the aspect ratio, extracted from panel (g).

Figure 1b and 1c present the simulated resonance frequency and polarization dependence, respectively, for graphene rectangle PPACs with a fixed aspect ratio ($AR = L_r/W_r$) of 4. Each array exhibits a distinct plasmonic resonance peak that redshifts as $L_r$ increases (Figure 1b and Figure S1a, Supporting Information). The resonance intensity also varies with $L_r$, which is primarily attributed to changes in the duty cycle of the array. The red curve in Figure 1b shows the fitted resonance trend based on the quasistatic coupled-mode theory (QCMT) model (Text ST1, Supporting Information).[30] Figure 1c illustrates the polar plot illustrating the dependence of resonance intensity on the polarization angle *θ*, clearly demonstrating a dipole-like resonance behavior. The resonance intensity reaches a maximum at *θ* = 0° and gradually decreases, vanishing completely as *θ* approaches 90°. At *θ* = 90°, the electric field component along the long axis of the PPAC becomes zero, thereby suppressing plasmon excitation. To quantitatively compare the polarization characteristics of the PPACs with different geometrical parameters, we defined the polarization ratio (*PR*) as,

$$PR = \frac{I_{max}}{I_{min}} \tag{1}$$

where $I_{max}$ and $I_{min}$ represent the maximum and minimum values of the absorption intensity at the resonance frequency, respectively. By substituting the absorption intensities of four representative graphene rectangle PPAC arrays with dimensions of 0.4 μm × 0.1 μm, 4 μm × 1 µm, 20 µm × 5 µm, and 80 µm × 20 µm into Equation (1), the calculated $PR$ are 154, 217, 211, and 127, respectively. These results indicate that the polarization dependence does not vary significantly in arrays with the same aspect ratio.

Furthermore, as shown in Figure 1e (also in Figure S1b, Supporting Information), the resonance frequency of graphene rectangle PPACs with varying aspect ratios (from 0.22 to 16) exhibit a pronounced redshift within the 0.1–4 THz range as $L_r$ increases from 1 µm to 80 µm. Notably, these PPACs also demonstrate clear dipole resonance according to their polarization behaviors (Figure 1f). The $PR$ values for rectangle PPACs with a fixed length ($L_r$ = 40 µm) and increasing widths ($W_r$ = 20–35 µm) are 11.1, 3.7, 1.8, and 1.2, respectively, demonstrating that a higher aspect ratio leads to a stronger polarization sensitivity.

In addition, while the resonance frequency associated with the long axis of the rectangle PPAC remains largely unaffected by the $\theta$ of the incident THz waves, the corresponding resonance intensity gradually diminishes as $\theta$ increases from 0° to 90° (Figure S2, Supporting Information). This behavior is characteristic of dipole plasmonic resonances, which are most efficiently excited when the incident electric field is aligned with the long axis of the PPAC. Interestingly, as $\theta$ increases, a second resonance mode gradually emerges at a higher frequency (Figure S2, Supporting Information), resulting in a clear saddle point in the polarization-dependent absorption spectra of each rectangle PPAC. This newly appearing resonance is attributed to plasmonic excitation along the width (short axis) of the PPAC. This assignment is further corroborated by FDTD simulations of the near-field distributions, which reveal strong localized fields oriented along the width direction at the higher-frequency resonance (Figure S2, insets, Supporting Information). This dual-resonance behavior—dependent on the excitation polarization—highlights a key advantage of the rectangle planar PPAC. In contrast to previous graphene−metamaterial integrated structures, which often require complex multi-layered architectures or anisotropic unit cell designs to support multiple polarization- or frequency-dependent responses,[5] the simple rectangular geometry of the PPAC achieves the same functionality. With a single, compact, and planar structure, the PPAC enables tunable and polarization-sensitive responses across a broad frequency range, significantly reducing design complexity and device footprint. Overall, compared with geometries of high symmetry, such as squares, circles, or regular hexagons, rectangle PPACs offer greater tunability due to their adjustable aspect ratio. This feature allows for more effective scaling in the low-THz frequency range without increasing the device footprint, while enabling a broader tuning range for both resonance frequency and polarization sensitivity.

Beyond the aspect ratio, the array spacing parameters $d_w$ and $d_l$ also influence the resonance frequency (Figure S3, Supporting Information). Specifically, a gradual redshift in resonance frequency is observed as $d_l$ decreases or $d_w$ increases, accompanied by an increase in resonance intensity as both $d_l$ and $d_w$ decrease. This behavior arises because the polarization direction of the incident THz wave is aligned parallel to $d_l$ direction. A reduction in $d_l$ narrows the spacing between adjacent cavities, thereby enhancing electromagnetic coupling. This increased coupling lowers the Coulomb restoring force associated with the collective oscillation of electrons in the cavity, leading to a decreased oscillation frequency and therefore a redshift of the resonance peak. Simultaneously, stronger near-field coupling traps more electromagnetic energy within the PPAC array, effectively reducing radiative losses and resulting in an enhanced absorption intensity.

To validate the above simulation results, we characterized the absorption spectra of the graphene rectangle PPACs using a THz time-domain spectrometer (THz-TDS, BATOP TDS–1008). To experimentally achieve sufficiently strong absorption signals, arrays of PPACs are prepared to enhance the overall THz absorption. To that end, PPAC arrays with unit dimensions of 50 μm × 10 μm, 40 μm × 10 μm, and 30 μm × 10 μm were fabricated on monolayer graphene transferred onto a high-resistivity $SiO_2$/Si substrate (see Methods), as shown in the scanning electron microscopy (SEM) images in Figure 1d. The graphene PPACs quality was further confirmed by confocal Raman spectroscopy (Methods and Figure S4a, Supporting Information). The Raman spectrum exhibits two prominent peaks at 1580 $cm^{-1}$ and 2700 $cm^{-1}$, corresponding to the G and 2D bands, respectively. The absence of a noticeable D peak and the symmetric single Lorentzian shape of the 2D peak—with a full width at half maximum of 32 $cm^{-1}$—indicate high-quality PPAC arrays. Subsequently, we measured the THz absorption spectra of the three PPAC arrays, with incident polarization parallel to the long axis of the PPACs. The absorption spectrum can be obtained as,

$$Abs = 1 - \frac{T_{PPAC}}{T_{SiO_2/Si}} \qquad (2)$$

where $T_{PPAC}$ is the transmission of the graphene PPAC arrays on the high-resistive $SiO_2$/Si substrate, and $T_{SiO_2/Si}$ is the transmission of the blank $SiO_2$/Si substrate. The resulting spectra, shown in Figure 1g, clearly demonstrate resonance features consistent with the simulation predictions. Moreover, all three rectangle PPAC arrays exhibit a redshift in resonance frequency with increasing aspect ratio (Figure 1h), which is also consistent with the simulation results.

## 3. Fabrication and characterizations of polarization-sensitive PPAC THz nanodetector

By combining the plasmonic resonance and the PTE effect within a rectangle PPAC and its array, sensitive and bias-free detection of incident THz waves at designated frequency and polarization states can be achieved.[4] To this end, we designed a THz nanodetector comprising source and drain electrodes bridged by an asymmetric graphene channel, as schematically illustrated in **Figure 2**a. The graphene channel was fabricated using monolayer graphene synthesized via chemical vapor deposition, transferred onto a high-resistivity $SiO_2$/Si substrate, and patterned using electron beam lithography (EBL). The left portion of the channel incorporates a graphene rectangle PPAC array interconnected by graphene nanoribbons, with individual resonators measuring 2.5 μm × 400 nm. In contrast, the right portion consists of an unpatterned graphene area (10 μm × 16 μm), resulting in a total channel area of 20 μm × 16 μm and a channel length ($L_c$) of 20 μm (Figure 2b). The absorption spectrum of the patterned plasmonic array exhibits a pronounced resonance at 2.52 THz within the 0.1–5 THz range (Figure S4b, Supporting Information), enabling efficient coupling with incident THz radiation. Source and drain electrodes were defined via ultraviolet lithography, followed by Ti/Au (10 nm/100 nm) deposition and lift-off. The current–voltage (*I–V*) characteristics of the detector under dark conditions (Figure S4c, Supporting Information) show a linear dependence, indicating excellent Ohmic contact between the electrodes and the graphene channel, with a measured resistance of 3852 Ω—an essential criterion for high-performance THz detection via the PTE effect.

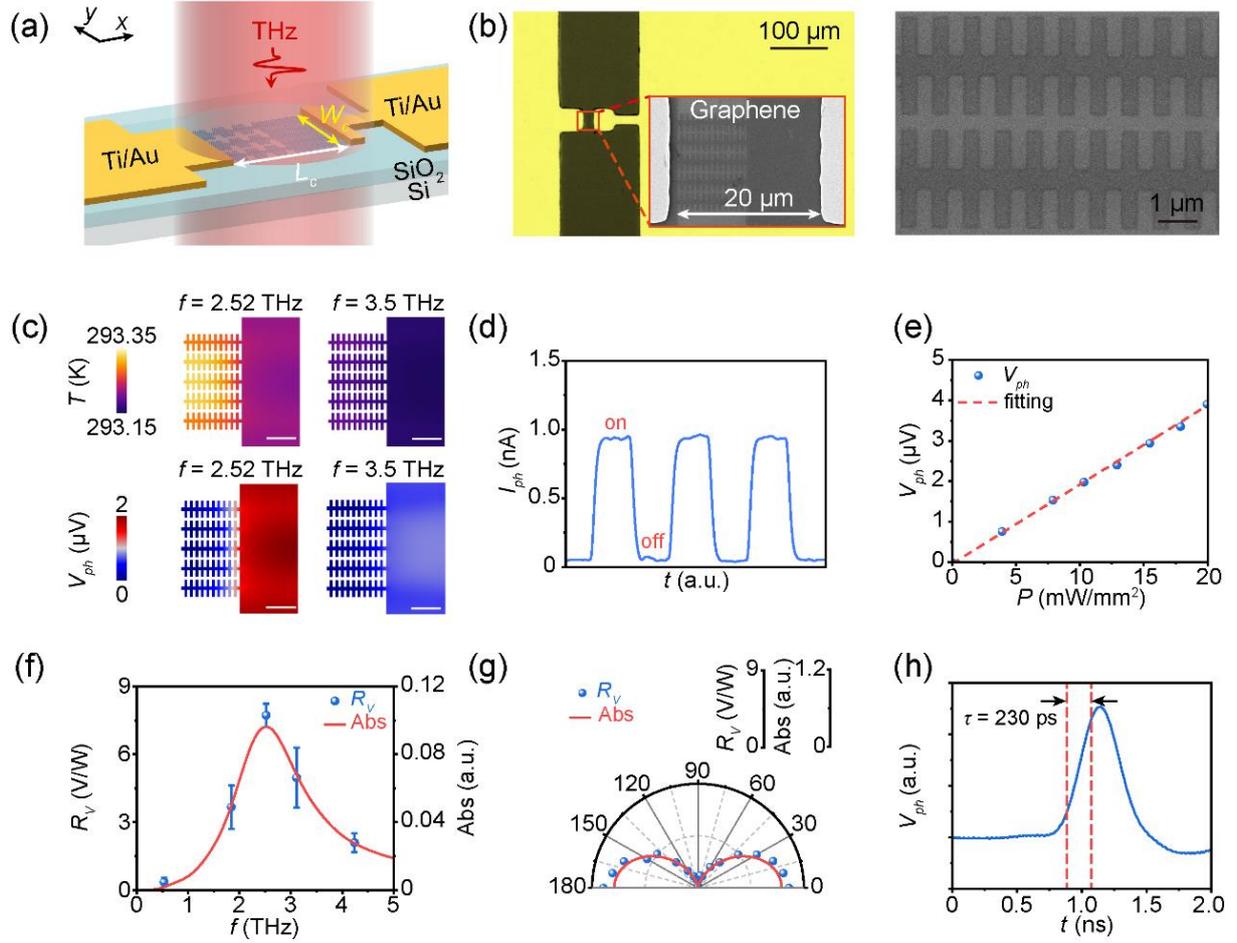

**Figure 2.** Fabrication and characterizations of the polarization-sensitive PPAC THz nanodetector. (a) Schematic illustration of the device, consisting of two Ti/Au metal electrodes connected by an asymmetric graphene channel. The asymmetric channel comprises a graphene rectangle PPAC array connected by graphene nanoribbons on one side and unpatterned graphene on the other. $L_c$ and $W_c$ denote the length and width of the graphene channel, respectively. (b) Left: optical microscope image of the fabricated device. Inset: SEM image showing the asymmetric graphene channel with a channel length of 20 μm. Right: SEM image of the graphene PPAC array. (c) Simulated temperature distribution (top panel) and electric potential distribution (bottom panel) of the device under THz illumination at the resonance frequency of 2.52 THz (left panel) and a non-resonance frequency of 3.5 THz (right panel). Scale bar: 5 μm. (d) Photocurrent response of the device under periodic 2.52 THz illumination. (e) Photovoltage as a function of illumination power. Dots: experimental data; dashed lines: linear fits. (f) Frequency-selective photoresponse (dots) and simulated THz absorption (solid line) of the device. (g) Polarization-dependent photoresponse (dots) and simulated polarization-sensitive THz absorption (solid line) of the detector. (h) Photovoltage of the device in response to a THz pulse (0.1–3 THz) illumination.

Before experimentally characterizing the device performance, we first conducted numerical simulations of several key parameters critical to the detection properties, thereby guiding subsequent experimental measurements. Specifically, we employed the finite element method (COMSOL Multiphysics) to simulate the electromagnetic field, temperature distribution, and electric potential of the device under both resonant and non-resonant excitation conditions (Figure S5, Supporting Information). Under THz wave irradiation, the photoresponse of the PPAC nanodetector emerges from two coupled phenomena: (i) PPAC-mediated field enhancement and hot-carrier generation, and (ii) asymmetric carrier/temperature gradients driving photovoltage formation at zero bias. First, SPPR modes in each PPAC unit of the cavity array strongly enhance electromagnetic field confinement, leading to significantly increased THz absorption (Figure S5a, Supporting Information) and the generation of a high-density hot-carrier population. Second, the asymmetric absorption profile across the graphene channel establishes a concentration gradient of these hot carriers. Additionally, hot-carrier relaxation can also induce a temperature gradient even under uniform illumination (Figure 2c, upper left panel). Driven by these gradients, the hot carriers diffuse from the PPAC array region toward the unpatterned graphene area, where they are ultimately collected by the electrodes. This process generates a measurable photovoltage $V_{\text{ph}}$ at zero bias (0 V), which can be quantified as,

$$V_{\text{ph}} = -\int_0^{L_c} S(x) \nabla T(x) dx \qquad (3)$$

$$S(x) = -\frac{\pi^2 k_B^2 T(x)}{3e} \left( \frac{d \ln \sigma}{d E} \right) \bigg|_{E=E_f} \qquad (4)$$

where $L_c$ is the length of the graphene channel, $S(x)$ is the Seebeck coefficient, $\nabla T(x)$ is the temperature gradient, $k_B$ is Boltzmann constant, $T(x)$ is absolute temperature, $e$ is the elementary charge and $\sigma$ is electrical conductivity of graphene. Two key observations can be made in the simulated temperature and electric potentials shown in Figure 2c (left panels): (i) the graphene PPAC array region maintains a higher temperature than the unpatterned region, and (ii) the thermal and carrier gradients drive a spontaneous current flow from the PPAC region to the unpatterned region without requiring any external bias voltage.

For comparison, the distributions of the electromagnetic field, temperature, and electric potential at a non-resonant frequency (e.g., 3.5 THz) were also simulated (Figure 2c, right panels; Figure S5b, right panel, Supporting Information). Evidently, the weak electromagnetic field localization in the graphene PPAC array at 3.5 THz significantly reduces THz absorption, leading to a diminished temperature gradient across the device and, consequently, a lower photovoltage. This frequency-dependent photovoltage behavior strongly indicates that photocurrent generation in the PPAC nanodetector is governed by the combined effects of SPPRs and the PTE effect, with the maximum photovoltage occurring at the plasmonic resonance frequency. Notably, the resonance absorption can be precisely tuned by tailoring the size of the PPAC unit and period of the PPAC array, thereby enabling the detector to exhibit both frequency- and polarization-sensitive photoresponses.

The THz response of the PPAC nanodetector was characterized under 2.52-THz illumination from a laser linearly polarized along the long axis of the rectangle PPAC unit. As shown in Figure 2d, significant photocurrent signals were generated at zero bias between the device electrodes. Three consecutive on/off cycles demonstrate excellent reproducibility of the photodetection response. Power-dependent measurements at 2.52 THz revealed a linear photocurrent−power relationship (Figure 2e), enabling calculation of the $R_V$ as,

$$R_V = \frac{V_{ph}}{P_{eff}} = \frac{I_{ph} \times R}{P_{eff}} \tag{5}$$

$$I_{ph} = \frac{2\pi\sqrt{2}V_{lock}}{4G} \tag{6}$$

where $V_{ph}$ and $I_{ph}$ are photovoltage and photocurrent upon a specific illumination power, respectively, $R$ is the resistance of the graphene channel, $V_{lock}$ is the photovoltage read out by the lock-in amplifier, and $G$ is the gain of the current preamplifier in V/A. Parameter $P_{eff}$ is the effective power. It is worth noting that the spot size of the THz beam (~1 mm diameter) is substantially larger than the active channel area of our antenna-free detectors. This contrasts with most conventional THz detectors integrated with antennas, where the device area is typically comparable to the beam size.[17,26,27] As a result, the incident power is not efficiently concentrated on the active region in our case, highlighting the importance of evaluating responsivity with respect to the effective power rather than the total incident power. Specifically, two approaches are employed to calculate the effective incident power $P_{eff}$.[18] The first approach defines $P_{eff}$ as the absorbed power within the active region of the device, expressed as,[4,18]

$$P_{eff} = \frac{P_0 \times S_{device}}{S_0} \times \Gamma_{abs} \tag{7}$$

where $S_{device}$ is the effective area (i.e., graphene channel area) of the device, $\Gamma_{abs}$ is absorption rate of PPAC array (Text ST1, Supporting Information), $P_0$ and $S_0$ are the power and spot size of the THz wave, respectively. This definition accounts for the fraction of incident power that is actually absorbed by the active region. The second approach instead considers the total incident power projected onto the active region, effectively removing the absorption rate $\Gamma_{abs}$ from Equation (7). This method assumes all incident power on the active area contributes to the response, regardless of the absorption efficiency. In our study, the PPAC detector exploits strong plasmonic resonances to achieve efficient near-field confinement and localized absorption of incoming THz waves within the graphene channel. This enables a substantial portion of the incident power to be effectively harvested, thereby justifying the use of absorbed power as a meaningful metric for responsivity evaluation. Moreover, this approach emphasizes the intrinsic potential of the PPAC architecture to surpass conventional limitations in THz detection sensitivity by leveraging resonant light–matter interactions at deeply subwavelength scales. The $R_V$ is calculated to be 7 V/W (0.7 V/W) at $L_c = 20$ μm referenced to absorbed power (incident power), which demonstrates that the detector maintains high responsivity despite its compact device area due to the incorporation of PPAC design (Table S1, Supporting Information).

The NEP and $D^*$ of the device were extracted from current noise density measurements (Figure S4d, Supporting Information). At low frequencies (1–1000 Hz), the noise spectrum is dominated by $1/f$ noise, while Johnson-Nyquist thermal noise prevails at higher frequencies (> 1000 Hz). The detector exhibits stable operation up to 7.8 kHz modulation frequency, as demonstrated by the consistent photocurrent response (Figure S4e, Supporting Information). Given this behavior, the NEP is primarily determined by Johnson-Nyquist thermal noise, which follows the expression,

$$\mathrm{NEP} = \frac{\sqrt{4k_B TR}}{R_V} \tag{8}$$

where $k_B$ is Boltzmann constant, with $T$ the room temperature. $D^*$ can be obtained as,

$$D^* = \frac{\sqrt{A}\sqrt{B}}{\mathrm{NEP}} \tag{9}$$

where $A$ represents the effective device area and $B$ the measurement bandwidth. Under 2.52-THz illumination, the nanodetector achieves an NEP of 1.2 nW/Hz$^{0.5}$ and a $D^*$ of $2.8 \times 10^5$ Jones, both are referenced to the absorbed power. Statistical analysis of four key parameters—$V_{ph}$, $R_V$, NEP, and $D^*$—was performed across multiple devices (Figure S6, Supporting Information), revealing their critical role in scaling to multi-pixel, large-area THz detector arrays. These parameters follow a normal distribution (linear scale), primarily due to variations in graphene quality (e.g., defects, impurities) between devices. The mean values are $V_{ph} = 0.9$ μV, $R_V = 4.8$ V/W, NEP = 4.1 nW/Hz$^{0.5}$, and $D^* = 2.6 \times 10^5$ Jones. Further improvements in $R_V$ and $D^*$, along with NEP reduction, can be achieved by optimizing graphene quality, fabrication processes, and device area, as discussed below.

To investigate the spectral response characteristics, we measured the photocurrent of the nanodetector under illumination at five representative THz frequencies: 0.53, 1.84, 2.52, 3.11, and 4.24 THz, and evaluated the corresponding $R_V$. The measured $R_V$ exhibits a clear frequency dependence that correlates well with the simulated THz absorption spectrum (Figure 2f). The strongest photoresponse is observed at 2.52 THz, which coincides with the plasmonic resonance frequency of the rectangle PPAC array. This result indicates that the device exhibits frequency-resolved photodetection behavior in the THz regime. Furthermore, by adjusting the size and period of rectangle resonator array, the photoresponse frequency of the device could be tuned over a broader THz region. To this end, we fabricated four detectors with varying rectangle dimensions (Figure S7, Supporting Information). *I–V* characteristics of these devices all show linear behavior with resistances on the order of several kΩ. In addition, each device displays a distinct resonance absorption peak (Figure S8, Supporting Information), indicating that the plasmonic resonance is highly tunable via geometric design. The responsivity of each device was assessed under illumination with different THz frequencies. As expected, the photoresponse varies across devices at a given frequency and reaches a maximum near their respective resonance conditions. For example, the device with PPAC sized at 2.5 μm × 400 nm exhibits a peak photocurrent at 2.52 THz (blue solid line), while a larger device with 30 μm × 20 μm rectangles shows peak response at 0.5 THz (red solid line). These findings highlight the capability of the graphene PPAC to support size-dependent plasmonic resonances, enabling frequency-selective photodetection across a broad THz spectral range. The flexibility in designing the PPAC geometry offers a powerful strategy for tailoring device operation toward specific THz frequencies, rendering it highly attractive for multi-band and tunable THz micro- and nano-detection applications.

The polarization-dependent resonance of the rectangle PPACs endows the nanodetector with polarization sensitivity—an essential feature for multi-parameter photodetection. To validate this mechanism, a half-wave plate was positioned in front of the device to systematically vary the polarization direction of the incident THz wave. The polarization angle *θ* was defined as 0° when the polarization direction was aligned along the longitudinal axis of a PPAC. As shown in Figure 2g, the responsivity of the detector exhibits a dipole-like dependence on the polarization angle, forming an inverted figure-eight pattern. This trend closely matches the simulated polarization-resolved THz absorption of the rectangle PPAC array, confirming that the polarization-sensitive capability of the nanodetector arises from the SPPR of the PPAC array. The polarization ratio of the detector, $PR_{device}$, can be quantified by,

$$PR_{\text{device}} = \frac{V_{\text{phmax}}}{V_{\text{phmin}}} \tag{10}$$

where $V_{\text{max}}$ and $V_{\text{min}}$ denote the maximum and minimum photovoltages for different illumination polarizations, respectively. Under 2.52-THz illumination, the $PR_{\text{device}}$ of the PPAC nanodetector is 93, which is outstanding among the current antenna/metamaterials-free THz detectors based on 2D materials (Table S1, Supporting Information). To exclude the potential influence of the source–drain electrodes on polarization sensitivity, we further analyzed their polarization response using FDTD simulations (Figure S9a, Supporting Information). The electromagnetic field distribution at the electrode surface varies with the polarization direction of the incident THz wave. Notably, the field distribution at the graphene–electrode interface exhibits a regular "8"-shaped polarization dependence, while the absorption behavior of the rectangle PPAC unit displays an inverted "8"-shaped pattern (Figure S9b, Supporting Information). The consistency between the photocurrent polarization dependence and the simulated polarization-resolved THz absorption confirms that the observed polarization sensitivity primarily originates from the SPPR of the rectangle PPAC array.

The response time $\tau$ is a critical parameter characterizing the performance of the PPAC nanodetector. The PTE mechanism, combined with the short channel of the detector and the high carrier mobility of graphene, enables high-speed detection performance. To characterize this, the temporal response of the rectangle PPAC nanodetector was excited and recorded simultaneously by an ultrafast THz pulse ranging from 0.1 to 3 THz (see Methods). It is important to note that the detection of ultrafast transient signals in the THz regime requires sufficiently strong photovoltage outputs of the detector to overcome system noise. To achieve this, multiple graphene channels were connected in series to form a 3-mm channel, enabling measurable signal levels with existing instrumentation (Figure S10, Supporting Information). As shown in Figure 2h, the induced photovoltage correlates directly with the incident THz pulse, revealing a response time of 230 ps, which surpasses most 2D material-based THz detectors (Table S1, Supporting Information). It is noted that the response time of the detector can be further reduced through various strategies like shortening channel lengths, implementing asymmetrical electrode designs, and enhancing the carrier mobility of graphene, which will be discussed later.[31,32]

## 4. Fabrication and characterizations of polarization-sensitive deep-subwavelength PPAC THz nanodetectors

The rectangle PPACs not only offer spectral tunability through an adjustable aspect ratio but also support polarization-sensitive and multi-frequency operation—without requiring additional structural modifications. These features make them highly promising for multifunctional, miniaturized, and integrable THz optoelectronic devices. To demonstrate this versatility, we experimentally fabricated a series of devices with graphene channel lengths ranging from 2 to 50 μm—approximately 1/60 to 1/2 of the incident THz wavelength—while keeping the channel width fixed at 16 μm (Figure S11a, Supporting Information). To complement the experimental findings and gain deeper insight into the underlying device physics, particularly the impact of channel length on THz detection performance, COMSOL Multiphysics simulations were carried out. Since the THz beam spot size covers both electrodes, the electrodes themselves can contribute to THz absorption and subsequent photocurrent generation. To account for this effect, the simulations modeled the THz response, temperature distribution, and electric potential profiles of the device under the combined influence of the metal electrodes and the graphene PPAC array.

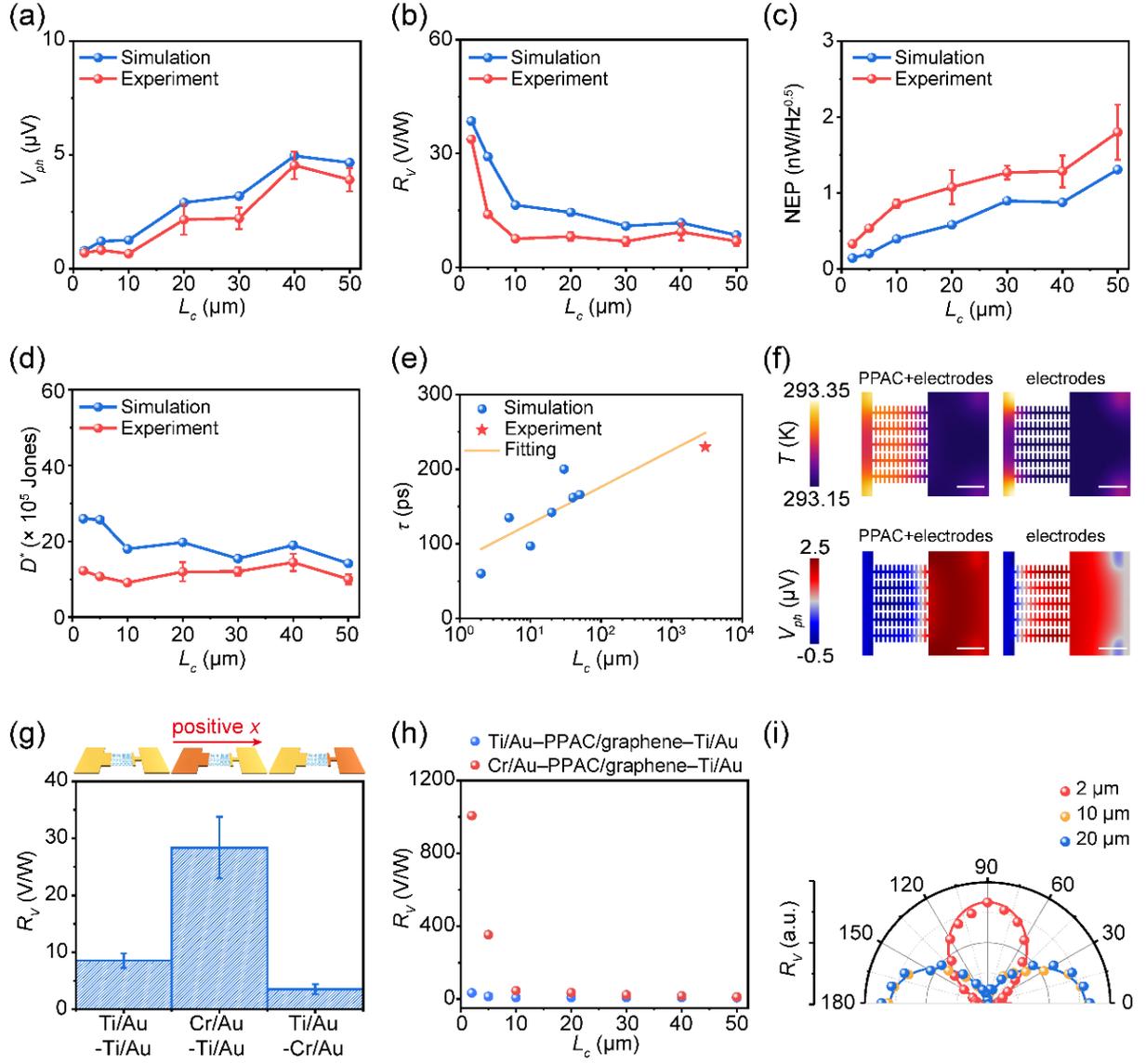

**Figure 3.** Graphene PPAC subwavelength THz detectors with different channel lengths. (a–d) Dependences of the photovoltage (a), $R_V$ (b), NEP (c), and $D^*$ (d) on channel lengths. Blue dots: simulated results. Red dots: experimental results. Lines: guides for the eye. (e) Dependence of response time $\tau$ on channel length. Blue dots are simulation results. Red star is experimental measurement from a 3 mm-length channel device. The orange line represents a linear fit of the simulated response time versus the logarithm of $L_c$. (f) Simulated temperature distributions (top panel) and electric potential distributions (bottom panel) in the detectors under two scenarios: the combined influence of the electrodes and graphene rectangle array (left panel), and the effect of the electrodes alone (right panel). The channel length of the detector is 20 μm. Scale bar: 5 μm. (g) Responsivity of detectors with different electrode configurations (left to right): Ti/Au−Ti/Au, Cr/Au−Ti/Au, and Ti/Au−Cr/Au. (h) Variation of the responsivity of Ti/Au–PPAC/graphene–Ti/Au and Cr/Au–PPAC/graphene–Ti/Au devices with different channel
16

length. (i) Polarization-dependent photoresponse (dots) and simulated polarization-sensitive THz absorption (solid line) for detectors with channel lengths of 2 μm (red), 10 μm (orange), and 20 μm (blue).

The measured resistances of the detectors increase from ~1800 Ω to ~8000 Ω as the channel length grows from 2 μm to 50 μm, in agreement with simulations where longer channels induce greater electron scattering and thus higher device resistance (Figure S12, Supporting Information). Under 2.52-THz illumination, both experimental measurements and numerical simulations reveal consistent photovoltage trends as a function of channel length (**Figure 3**a). Two distinct behaviors are observed: (i) the photovoltage remains nearly constant in the short-channel regime ($L_c < 10$ μm); and (ii) for channel lengths exceeding 10 μm, the photovoltage increases and gradually saturates for $L_c > 20$ μm.

To elucidate the underlying mechanisms across different channel lengths, we conducted simulations of the temperature field, electric potential distribution, and resulting photovoltage under two configurations: (i) electrodes only, and (ii) electrodes combined with the PPAC array. In the electrode-only configuration, heat accumulation is confined near the contact regions, with potential drops localized around the electrodes (Figure 3f, right panel). In contrast, incorporating the PPAC array leads to elevated temperatures within the cavity region and extends the potential variation into the central channel, resulting in a higher overall device potential (Figure 3f, left panel). Importantly, the photovoltage difference between these two configurations increases with channel length (Figure S13a, Supporting Information), highlighting the growing contribution of the PPAC array to THz response in longer channels. By subtracting the electrode-only photovoltage from the combined system, we isolate the PPAC-induced component (Figure S13b, Supporting Information), which is negligible for short channels ($L_c < 10$ μm). In this regime, the dominant contribution arises from strong electromagnetic field confinement between closely spaced electrodes, as confirmed by near-field distributions (Figure S14, Supporting Information). These results demonstrate that in short-channel devices, the photoresponse is governed primarily by electrode-enhanced field concentration, rather than plasmonic resonance effects in the PPAC array—accounting for the observed plateau in photovoltage for $L_c$ smaller than 10 μm. As the channel length increases, the role of the PPAC array becomes more prominent, enabling efficient photovoltage generation via localized plasmon polariton resonance-induced PTE effect.

For devices with channel lengths $L_c > 10$ μm, the photovoltage response is primarily governed by the diffusion dynamics of thermal energy generated by PPACs. When the cumulative length of the PPAC array remains below the thermal diffusion length, which is calculated as 15.9 μm (see Text ST2, Supporting Information for details), extending the array, *i.e.*, incorporating additional PPACs, leads to a collective increase in local heating, particularly at the leftmost PPAC–electrode interface. This thermal accumulation amplifies the lateral temperature gradient across the device, thereby enhancing the thermoelectric photovoltage (Figure 3f, and see Text ST2 and Figure S15a in Supporting Information for detailed discussion). In contrast, when the PPAC array extends beyond the thermal diffusion length, the incremental contribution of newly added PPACs becomes negligible. These additional units no longer influence the temperature at the electrode interface, resulting in a stabilized temperature gradient and thus a saturated photovoltage output (Figures S15b and S15c, Supporting Information). Accordingly, the photovoltage increases monotonically with channel length until it asymptotically saturates in long-channel devices ($L_c > 20$ μm).

These observations reveal that the photovoltage generation in PPAC nanodetectors with channel lengths in the range $L_c < 50$ μm arises from two distinct physical mechanisms: (i) localized electromagnetic field enhancement at the metal electrodes, and (ii) SPPR-induced photothermal effects in the graphene PPAC arrays under THz excitation. Specifically, in short-channel devices ($L_c < 10$ μm), the photovoltage is dominated by strong THz near-field confinement and capacitive coupling at the electrode edges, while in long-channel devices ($L_c > 10$ μm), it is primarily contributed by plasmonic resonant energy absorption and thermal diffusion processes within the PPAC array. This dual-regime behavior underscores the interplay between electrode-induced field concentration and SPPR-enabled PTE conversion in shaping the photoresponse of the nanodetector.

Under a constant incident THz beam spot size, the effective absorbed power $P_{eff}$ decreases significantly—by a factor of 30 (Figure S11b, Supporting Information)—as the channel length is reduced from 50 μm to 2 μm. In contrast, the photovoltage drops more gradually, by a factor of only 6 (Figure 3a). Notably, for devices with $L_c < 10$ μm, the photovoltage remains nearly constant despite the shrinking channel length. This disparity leads to a clear enhancement in the device responsivity $R_V$, which increases by a factor of 4.9 as the channel length narrows from 50 μm to 2 μm (Figure 3b). Consequently, the NEP, which is inversely proportional to $R_V$, decreases significantly by a factor of 5.5 over the same range (Figure 3c). Meanwhile, the specific detectivity $D^*$ shows a more modest increase, improving by a factor of 1.2 (Figure 3d). These improvements become especially pronounced when the channel length falls below 10 μm. At a minimum channel length of 2 μm, the device reaches its peak performance, with a maximum responsivity of 34 V/W and a minimum NEP of 0.3 nW/Hz$^{0.5}$, all within a compact active area of just 32 μm$^2$, corresponding to a deep-subwavelength scale of ~$\lambda_0$/60.

In addition to enhanced sensitivity, the response time also decreases significantly as the channel length shortens. However, for shorter channel lengths, individual devices generate substantially smaller voltage outputs, resulting in signal amplitudes that are too weak for reliable time-domain detection with our current setup. To address this, we simulate the response time of detectors with channel lengths reduced from 50 to 2 μm, observing an approximately linear dependence on the logarithm of the channel length, $L_c$ (Figure 3e and Text ST3 in the Supporting Information for details on response time extraction). At a channel length of 2 μm, the simulated response time reaches ~60 ps, representing a 2.8-fold reduction compared to the 50 μm case. This trend can be attributed to the combined effects of high carrier mobility in graphene and the PTE mechanism. Shorter channels facilitate faster electron–electron thermal equilibrium and reduce the characteristic carrier scattering time, thereby accelerating the dynamic response of the device. Importantly, the experimentally obtained response time for the 3-mm channel device falls on the extrapolated curve of the theoretical simulation, further validating the model and confirming its applicability to shorter channel lengths. These findings suggest that scaling down the device not only improves responsivity and integration density but also enables ultrafast detection speeds—an essential advantage for high-performance, miniaturized THz optoelectronic systems.

To further enhance detector responsivity, we introduced electrodes with asymmetric work functions. Specifically, we fabricated PPAC nanodetectors with Ti/Au–PPAC/graphene–Ti/Au, Cr/Au–PPAC/graphene–Ti/Au, and Ti/Au–PPAC/graphene–Cr/Au configurations, all resonant at 2.52 THz (see schematic and SEM images in Figure S16, Supporting Information). Each device features a 20 μm × 16 μm channel and individual PPAC measuring 400 nm × 2.5 μm. The metal electrodes consist of 10 nm Ti or Cr, capped with 100 nm Au (Figure 3g, top panel). Photocurrents were measured under 2.52-THz laser excitation at zero bias, revealing that the Cr/Au–PPAC/graphene–Ti/Au device exhibited the highest responsivity, while the Ti/Au–PPAC/graphene–Cr/Au structure showed the lowest (Figure 3g). This asymmetry-dependent behavior underscores the critical role of electrode work function engineering in optimizing THz photoresponse.

We further analyzed the operation mechanisms of the three devices, as illustrated in the bottom panels of Figure S16 (Supporting Information). Importantly, both the asymmetry of the graphene channel and the asymmetry of the metal electrodes must be considered to fully understand the photocurrent generation. On one hand, the asymmetric graphene channel, incorporating the rectangle PPAC array, supports SPPRs, which elevates the hot-carrier temperature on the left side of the device. This creates a spatial temperature distribution $T(x)$ across the channel, driving hot carriers from the high-temperature region to the low-temperature region. As a result, the thermoelectric photocurrent $I_0$ flows along the positive $x$-axis in all devices, consistent with the $p$-type doping of graphene under ambient conditions.[33] On the other hand, the electrodes with different work functions (Ti and Cr) induce distinct levels of metal-induced doping in graphene at the two terminals, resulting in a nonuniform local Fermi level $E_f(x)$. This leads to a position-dependent Seebeck coefficient $S(x)$ and built-in potential gradient $\nabla V(x)$, as described by Equations (3) and (4). For $p$-type graphene with a positive $S(x)$, the work function differences ($W_{Gr} > W_{Cr} > W_{Ti}$) cause limited electron transfer from Cr/Au electrode to graphene, but substantial electron transfer from Ti/Au electrode to graphene.[34–36] Consequently, in the Cr/Au–PPAC/graphene–Ti/Au device, the electrode-induced photocurrent $I_e$ flows along the positive $x$-axis, reinforcing the SPPR-induced hot-carrier current $I_0$ (Figure S16b, bottom panel, Supporting Information). In contrast, in the Ti/Au–PPAC/graphene–Cr/Au device, $I_e$ flows in the opposite direction, partially canceling $I_0$ (Figure S16c, bottom panel, Supporting Information). This interplay explains the stronger photocurrent observed in the Cr/Au–PPAC/graphene–Ti/Au device, in agreement with experimental measurements.

The photocurrent characteristics of the Cr/Au–PPAC/graphene–Ti/Au device with a 20 μm channel length were further examined under 2.52-THz irradiation at varying powers. The photocurrent increases linearly with incident power (Figure S17a, Supporting Information). Frequency-dependent measurements reveal a $R_V$ peak of 35 V/W at 2.52 THz, consistent with the absorption maximum of the PPAC array (Figure S17b, Supporting Information). Compared to the symmetric Ti/Au–PPAC/graphene–Ti/Au configuration, the Cr/Au–PPAC/graphene–Ti/Au device exhibits a more pronounced enhancement in responsivity as the channel length decreases—attributable to the stronger PTE effect arising from the work function difference between the asymmetric electrodes (Figure 3h). Specifically, the regions of metal-induced doping in graphene at two terminals—governed by work function difference between asymmetric electrodes and graphene—remains fixed, with only hot carriers confined within these regions exhibiting significantly enhanced PTE contributions. As the channel length increases from 2 μm to 50 μm, the metal-induced doped regions constitute a diminishing fraction of the total channel, leading to a less significant contribution of this enhancement mechanism in devices with $L_c > 20$ μm. This result highlights the critical role of introducing asymmetric electrodes to enhance built-in potential gradients and boost PTE-driven photocurrent generation. Notably, for a detector with an area of 2 μm × 16 μm, the Cr/Au–PPAC/graphene–Ti/Au architecture achieves a responsivity of 1007 V/W, an NEP of 16 pW/Hz$^{0.5}$, and a $D^*$ of $2.9 \times 10^7$ Jones—exceeding the symmetric device by more than one order of magnitude.

To evaluate the polarization sensitivity of the detector, we measured the photocurrent response as a function of the polarization angle for various channel lengths (Figure 3i). For devices with $L_c$ = 10 μm and 20 μm, the response exhibits an inverted "8"-shaped dipolar polar plot pattern, consistent with the polarization-selective absorption characteristics of the PPAC array (Figure 1c and Figure S9b, Supporting Information). The polarization ratios for these two detectors are 32 and 93, respectively. In contrast, for the short-channel device with $L_c$ = 2 μm, the polarization response reverses to a regular "8"-shaped dipolar distribution, with a polarization ratio of 9. This inversion originates from the dominant influence of electrode-induced electromagnetic field localization, which overtakes the polarization-dependent response of the PPAC array at shorter channels (Figure S9b, Supporting Information). This behavior confirms that in short-channel devices, the photovoltage generation is primarily governed by the near-field enhancement and asymmetry introduced by the electrodes. Conversely, in long-channel devices, the response is dominated by the SPPRs of the PPAC array. Notably, devices with reduced channel lengths retain polarization sensitivity while introducing an additional degree of freedom in detector design—arising from the tunable competition between electrode-induced and PPAC-mediated effects.

These results demonstrate that, by reducing the channel length and introducing asymmetric electrodes, one can simultaneously enhance the local THz near-field intensity and leverage both the work function difference and thermal conductivity mismatch between the source and drain. This synergistic effect amplifies the temperature gradient across the graphene channel, thereby substantially boosting the responsivity, while maintaining the monolithic frequency and polarization selectivity.

## 5. Performance comparison and THz imaging of PPAC nanodetector

In contrast to most 2D-material-based THz detectors that rely on external antennas or metamaterials for THz wave coupling, the PPAC design operates antenna-free, achieving an exceptionally small minimum channel length of 2 μm—just 1/60 of the incident THz wavelength. A comparison with existing antenna-free 2D THz detectors clearly demonstrates that, despite its compact footprint, the PPAC nanodetector delivers superior performance, with a responsivity of 98 V/W and an NEP of 165 pW/Hz$^{0.5}$, both referenced to the incident power (**Figures 4**a and 4b).[4,5,17,18,20,25,26,32,37–56] It is worth noting that, compared to our previously reported disk-shaped PPAC detector,[4] the rectangle PPAC nanodetector achieves over two orders of magnitude reduction in device area while boosting responsivity by more than one order of magnitude. This enhancement arises from the synergistic effect of the rectangle PPACs and the asymmetric metal contacts within the short-channel device, which together induce significantly stronger electromagnetic field confinement than the disk-shaped PPACs (Figure S18, Supporting Information). Notably, the $R_V$ and NEP of the PPAC nanodetector, referenced to the absorbed power—1007 V/W and 16 pW/Hz$^{0.5}$, respectively—are both competitive with state-of-the-art room-temperature antenna-coupled 2D THz detectors, and even surpass many of them (Figures 4a and 4b). What truly sets this device apart from existing THz detectors, however, is its unique combination of deep-subwavelength footprint, polarization sensitivity, and high-speed performance (Table S1, Supporting Information).

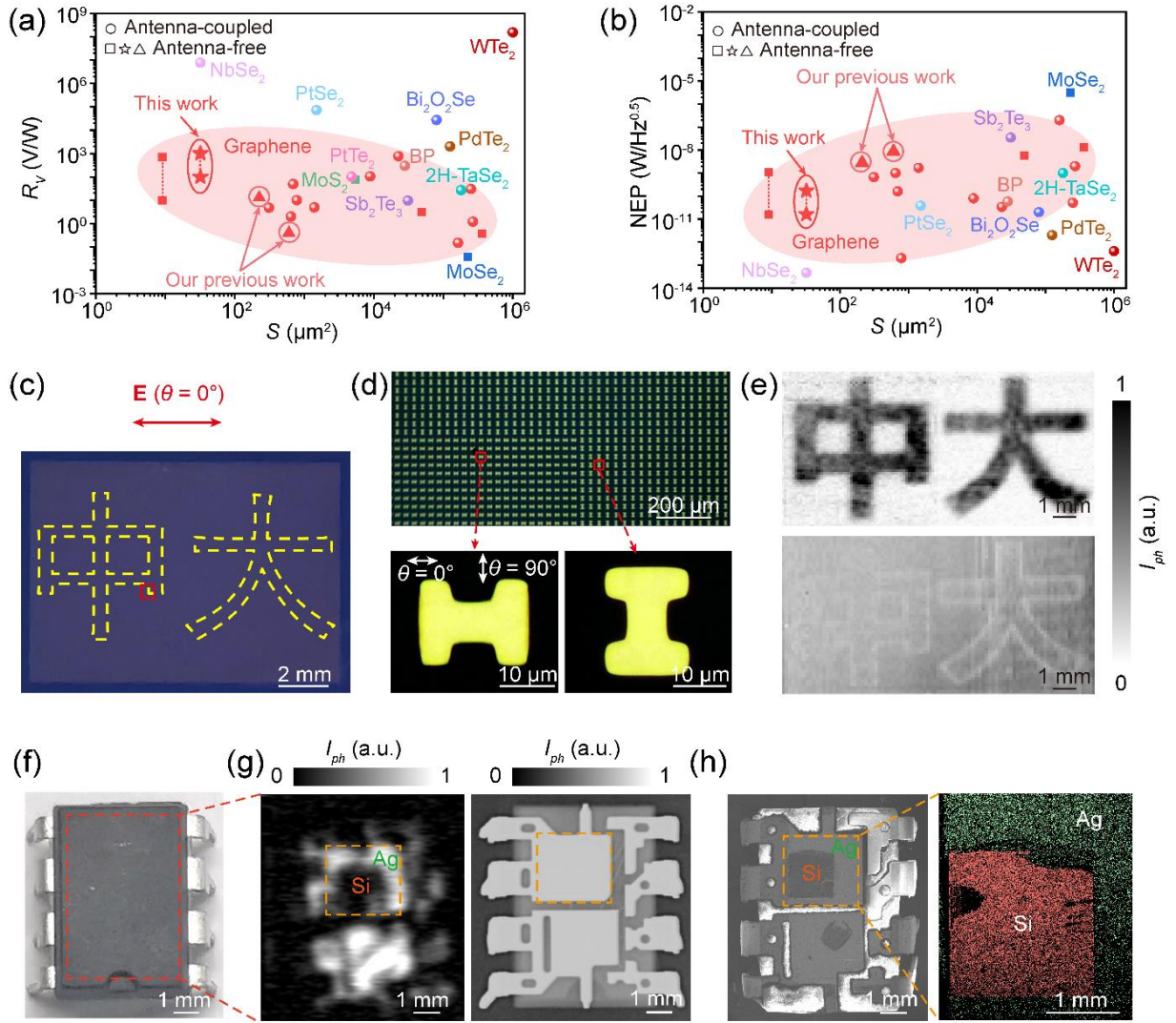

**Figure 4.** THz detection performance comparison and THz imaging of the PPAC nanodetector. (a, b) Comparison of the $R_V$ (a) and NEP (b) of the PPAC nanodetector with other room-temperature 2D-materials-based THz detectors. (c) Photograph of the imaging object with two

Chinese characters, "中" and "大", where the outer and inner regions of the two characters are filled with different metal microstructure arrays. (d) Optical microscope image of metal microstructure arrays (top panel). Region enclosed by red box shown in (d) is filled with H-shaped (bottom left panel) and inverted H-shaped (bottom right panel) microstructures. (e) Photocurrent imaging of the two Chinese characters with PPAC nanodetector (top panel) and a commercial detector (bottom panel) at 2.52 THz. (f) Photograph of the power management integrated circuit (PMIC). (g) Comparative imaging of the internal structure of the PMIC covered by the cap layer using stealth THz imaging with the PPAC detector (left) and conventional X-ray imaging (right). The THz image, highlighted by the dashed orange square, clearly differentiates regions of varying conductivity, such as silicon and metal, which cannot be distinguished in the corresponding X-ray image. (h) SEM image (left) and energy-dispersive X-ray spectroscopy (EDS) mapping (right) of the internal structure of the PMIC after removal of the top cap layer. The EDS mapping reveals the elemental distribution, which correlates well with the conductivity contrast identified in the THz image shown in panel (g).

The photoresponse of the PPAC nanodetector enables non-destructive, non-contact imaging in the THz regime. In particular, polarization-resolved THz imaging provides critical insights into anisotropic materials, allowing extraction of material-specific information that is inaccessible through intensity-only measurements. As a proof of concept, we employed the device within a polarization-resolved dual-focus scanning imaging system to visualize samples with engineered internal anisotropic orientations (Figure S19, Supporting Information). The sample was positioned on a 2D electronically controlled translation stage, located at the focal plane of the first parabolic mirror, while the detector was placed at the focal plane of the third mirror. A 2.52-THz linearly polarized laser, with the outpout frequency matching with the peak response frequency of the detector, was used as the light source. A quarter-wave plate was placed in front of the sample to convert linear polarization into circular polarization, thereby eliminating polarization dependence from the laser itself. By raster scanning the sample and recording the photocurrent generated by THz waves transmitted through the object and incident on the detector, a 2D image of the hidden feature is reconstructed (see Methods for details). The imaging targets were two Chinese characters, "中" (meaning "middle") and "大" (meaning "large"), patterned using metallic microstructure arrays with identical geometry and dimensions but orthogonal orientations (Figure 4c). The inner regions of both characters were filled with "H"-shaped microstructure arrays, while the surrounding areas were filled with inverted "H"- shaped arrays (Figure 4d). Simulated transmission spectra show that both the "H"-shaped microstructure arrays and the PPAC nanodetector share the same resonance frequency and polarization selectivity (Figure S20, Supporting Information). Specifically, the transmittance of the microstructure array at 2.52 THz is nearly zero at a polarization angle $\theta = 0°$, but reaches ~70% at $\theta = 90°$. As a result, when circularly polarized light passes through the object, the transmitted polarization state differs between the character regions and their surroundings—allowing the nanodetector to discern them. The polarization-sensitive image acquired at 2.52 THz using the PPAC nanodetector (Figure 4e, top panel) consists of 90 × 50 pixels with a step size of 0.2 mm × 0.2 mm. It clearly resolves the "中" and "大" characters, confirming that the detector can differentiate orthogonal polarization states and extract embedded polarization information, thereby enhancing image contrast and resolution. For comparison, the same object was imaged using a commercial polarization-insensitive THz detector (Golay cell, Thomas Keating Instruments, TK100), shown in bottom panel of Figure 4e. The Golay cell fails to resolve the inner character regions, capturing only faint contours, which are due to boundary scattering. This comparison confirms that polarization-insensitive detectors cannot capture rich polarization information, underscoring the unique capability of the PPAC nanodetector.

Importantly, the polarization-resolved capability of the PPAC nanodetector in the THz regime enables the detection and identification of concealed internal structures, highlighting its potential for non-destructive inspection of semiconductor chips. The internal architecture of a chip, typically protected by a thin cap layer, consists of heterogeneous materials with distinct carrier concentrations, such as metallic Ag and semiconducting silicon. These variations induce differences in both the intensity and polarization state of THz waves reflected from different regions of the chip.[57] When a linearly polarized THz wave is reflected from surfaces with different carrier concentrations, its polarization state is altered to varying degrees, leading to corresponding changes in the intensity of the two orthogonal electric field components. In this context, we define the horizontal direction as parallel to the long axis of the rectangle PPAC unit cell (Figure S21, Supporting Information), which is the sensitive polarization direction of the detector. Because the PPAC detector selectively responds to the horizontal component of the THz wave, variations in reflected intensity along this direction can be directly exploited to distinguish between materials. To validate this mechanism, Fresnel-equation-based simulations were carried out at 2.52 THz to compare the reflected polarization states from Ag and silicon (Figure S21, Supporting Information). The results show that Ag behaves as a nearly perfect reflector, whereas silicon with reduced charge carriers introduces a measurable polarization change, resulting in a weaker horizontal component relative to Ag. Such differences can be sensitively detected by the PPAC polarization-resolved detector, thereby enabling precise, non-destructive evaluation of the internal structure of semiconductor chips.

As a practical demonstration, we imaged the internal structure of a power management integrated circuit (PMIC, F17D39) covered by a thin cap layer (Figure 4f) using the reflective dual-focus scanning imaging system (Figure S22, Supporting Information). A linearly polarized 2.52 THz laser beam, incident at 45°, was focused onto the PMIC mounted on a 2D electronically controlled translation stage, while the reflected THz signal was recorded by the PPAC nanodetector. The resulting THz image (Figure 4g, left panel) reveals distinct contrast between different regions within the chip, enabling their clear identification. To further verify these regions, the protective cap layer was removed, and the internal structure was characterized using optical photography, SEM, and energy-dispersive X-ray spectroscopy (EDS). The results confirm that the PMIC consists of a silicon chip mounted on an Ag base (Figure 4h and Figure S23, Supporting Information), consistent with the contrast observed in the THz image. This demonstrates that THz polarization imaging can effectively distinguish Ag and silicon regions with different carrier concentrations inside the chip. For comparison, the same chip was analyzed using polarization-insensitive X-ray transmission imaging (VJ Technologies, IXS080BP210P396) (Figure 4g, right panel). Unlike the THz image, the X-ray result shows nearly identical grayscale values for the Ag and silicon regions, providing insufficient contrast for reliable material discrimination. This comparison underscores the unique advantage of the PPAC nanodetector in capturing both intensity and polarization information—capabilities not accessible with conventional X-ray imaging—highlighting its promise for non-destructive chip inspection and material identification.

## 6. Conclusion

In summary, we have realized a miniaturized, polarization-sensitive, and frequency-selective PPAC nanodetector that exploits plasmon polariton resonance modes in graphene rectangle PPACs integrated with asymmetric metal electrodes. This self-powered PPAC nanodetector operates without external THz antennas and achieves a deep-subwavelength channel length of only 2 μm—approximately 1/60 of the incident wavelength. Despite its compact, antenna-free architecture, one device configuration demonstrated state-of-the-art performance for room-temperature 2D THz detection, with a high responsivity of 1007 V/W, a low NEP of 16 pW/Hz$^{0.5}$, a detectivity of $2.9 \times 10^7$ Jones, and a theoretically achievable response time down to 60 ps. Complementary polarization-oriented designs further revealed the capability to harness the frequency-selective plasmonic response of rectangular PPACs, enabling broadband operation from 0.53 to 4.24 THz together with an exceptionally high polarization ratio of 93. These attributes collectively allow non-destructive imaging of concealed semiconductor chips and polarization-resolved characterization of anisotropic materials. Looking forward, scaling this architecture into detector arrays offers a pathway toward larger photosensitive areas, enhanced sensitivity, and multi-parameter focal-plane imaging. The demonstrated PPAC nanodetector thus establishes a powerful platform for high-resolution THz radar, spectroscopy, and semiconductor inspection, advancing the frontiers of miniaturized, polarization- and frequency-sensitive THz nanophotonic devices.

## 7. Methods

*Graphene transfer and device fabrication:* The high-resistive $SiO_2$/Si substrate, with a Si thickness of 500 μm, a $SiO_2$ layer thickness of 300 nm and a resistivity greater than 20000 Ω·cm (Nanjing MKNANO Tech. Co., Ltd. www.mukenano.com), exhibits high transparency and insulation properties in THz region.[37] A monolayer graphene film (Jiangsu XFNANO Materials Tech. Co., Ltd.), coated with polymethyl methacrylate (PMMA) as a protective mask, was transferred onto the $SiO_2$/Si substrate via a wet transfer technique. Afterward, the sample was immersed in acetone for one hour to remove the PMMA. The monolayer graphene was patterned using EBL and then etched by oxygen plasma ($O_2$ at 400 Pa, power of 18 W, etching for 2 min) to create a asymmetric graphene channel with individual rectangle PPAC size of 2.5 μm × 400 nm. Subsequently, the electrodes were defined using ultraviolet maskless lithography machine (TuoTuo Technology, UV Litho-ACA) and coated with Ti/Au or Cr/Au (ZhongNuo Advanced Material (Beijing) Technology Co., Ltd.) in 10 nm/100 nm thicknesses through thermal evaporation. To remove the organic residues, the PPAC nanodetector was annealed in a tube furnace with an Ar/$H_2$ mixture flow of 150 sccm/350 sccm at 450 °C for 90 min. The device was then affixed to a printed circuit board (PCB) and connected to PCB pads by aluminum wire bonding for further photoresponse characterizations.

*Characterizations:* The morphology and Raman spectrum of graphene were characterized using a SEM (FEI, Quanta 450) and a confocal Raman spectrometer (Renishaw, Invia Reflex), respectively. The elements of the internal structure of the chip were characterized using an EDS (Oxford Instruments, 51-XMX020). The transmission spectra of graphene rectangle PPACs were tested using a THz time domain spectroscopy (BATOP, TDS-1008). The electrical transport characteristics of the PPAC nanodetector were measured using a source meter (Tektronix, Keithley 2636b). A far-infrared gas laser (Edinburgh Instruments, FIRL 100) with output frequencies of 4.24 THz, 3.11 THz, 2.52 THz, 1.84 THz and 0.53 THz was employed for THz detection. The modulation frequency of the laser was controlled by an optical chopper (SCITEC, 310CD). A low-noise current pre-amplifier (FEMTO, DLPCA-200) was employed to amplify the signals from the PPAC nanodetector, which were subsequently processed using a lock-in amplifier (Stanford, SR830) and a digital phosphor oscilloscope (Tektronix, DPO7354C) for readout. To assess the noise current spectrum of the PPAC nanodetector, a semiconductor parameter analyzer (PDA FS-Pro) was utilized to assess the noise current spectrum of the device.

*Photoresponse measurements:* In the photoresponse measurement, a laser beam (~ 1 mm spot diameter), modulated by an optical chopper at 225 Hz, was fully irradiated on the device. Fine-tunning was carried out to optimize the photocurrent. The generated photocurrent was then converted to a voltage signal using a current pre-amplifier and recorded by a lock-in amplifier. For 2D imaging of a specific object, the transmitted light from the imaging object, illuminated by a laser beam modulated at 225 Hz, was directed onto the device to produce photocurrent. The object was scanned using a 2D electrically controlled displacement stage. The photocurrents at each position of the object were measured by a current pre-amplifier and recorded by a lock-in amplifier. By correlating the photocurrent distribution with the scanning positions, a 2D image of the object was reconstructed.

The response time of the PPAC detector was evaluated using THz pulses exhibiting a peak power of 2 mW and a pulse energy of 1.4 μJ. These THz pulses were generated by pumping a LiNbO$_3$ crystal with an 800-nm femtosecond laser (Spectra-Physics, Solstice Ace) and guided by off-axis parabolic mirrors to irradiate onto the PPAC detector. The photovoltage signal generated by the detector, denoted as $R(t)$, underwent amplification via a low-noise preamplifier (FEMTO, DLPCA-200) with a gain of G = $10^3$ before acquisition by a high-bandwidth oscilloscope (Tektronix DPO7354C, 3.5 GHz bandwidth). The voltage signal $V(t)$ captured by the oscilloscope represents the amplified photovoltage signal, incorporating the temporal delay $M(t)$ introduced by the preamplifier. Consequently, the true photovoltage $R(t)$ was obtained by deconvolving the observed voltage signal $V(t)$ with the delay function $M(t)$ of the preamplifier, expressed as $V(t) = \int_0^t R(t')M(t-t')dt'$. To recover the real photovoltage response of the PPAC detector, a high-speed near-infrared probe was used to calibrate and obtain $F[M(t)] = F[V_{IR}(t)]/F[R_{IR}(t)]$, where F denotes the Fourier transform. The actual THz photovoltage response $R_{THz}(t)$ of the PPAC detector to THz radiation was then determined by deconvolving the observed THz voltage signal $V_{THz}(t)$ with the delay function $M(t)$ of the preamplifier, defined as $V_{THz}(t) = \int_0^t R_{THz}(t')M(t-t')dt'$.

*Simulation:* The electromagnetic field, temperature field and electric potential distributions of the PPAC nanodetector were calculated using COMSOL. The absorption spectrum of graphene rectangle PPAC array was simulated using FDTD. First, the electromagnetic field distribution was determined using the "Electromagnetic Waves, Frequency Domain (EWFD)" module. The graphene's optical conductivity in the THz region can be defined as,

$$\sigma = \frac{je^2 E_f}{\pi \hbar^2 (\omega + j\tau^{-1})} \tag{11}$$

where Fermi level $E_f$ and scattering rate ($\hbar\Gamma = \hbar\tau^{-1}/2$) were set to -0.3 eV and 0.003 eV, respectively. The current density can be expressed as $\boldsymbol{J} = \sigma \boldsymbol{E}$.[58] The refractive indices of the Si and SiO$_2$ substrates were set to 3.42 and 1.955, respectively.[59,60] In temperature field distribution simulation, the energy dissipation (heat source) distribution, $q_{absorbed} = 0.5 Re(\boldsymbol{J} \cdot \boldsymbol{E}^*)$, calculated from the EWFD module, was input into the "Heat Transfer in Solids (HT)" interface to calculate the temperature distribution, where graphene's thermal conductivity was set to 5300 W/(m·K).[61,62] The temperature of graphene PPAC nanodetector can be obtained by,

$$C_p \frac{dT}{dt} = -\nabla \cdot q + Q \tag{12}$$

$$q = -\kappa \nabla T \tag{13}$$

where $C_p$ and $\kappa$ are the heat capacity and thermal conductivity of graphene, respectivily, and $Q$ is heat source. For the d$T$/d$t$ term in the equation, it equals zero under steady-state conditions and is non-zero during transient conditions. Since the process of the heat transfer in graphene is primarily mediated by hot carriers, the heat capacity should be set as the carrier heat capacity, which is significantly smaller than the lattice heat capacity (~700 J/(kg·K)). Typically, the heat capacity of hot carriers in graphene is 10$^{-2}$ to 10$^{-4}$ times that of the lattice heat capacity,[63] which was set to 10 J/(kg·K).

The "Electric Currents" interface was used to calculate the electric potential $V$ of the PPAC nanodetector, which can be obtained as,

$$\nabla \cdot \boldsymbol{J} = 0 \tag{14}$$

$$\boldsymbol{J} = \sigma_d \boldsymbol{E_d} \tag{15}$$

$$\boldsymbol{E_d} = -\nabla V \tag{16}$$

where $\boldsymbol{J}$ is the current density, $\sigma_d$ and $\boldsymbol{E_d}$ represent the DC electrical conductivity and DC electric field, respectively. The "Electric Currents" and "Heat Transfer in Solids" interfaces were coupled via the "Thermoelectric Effect" and "Electromagnetic Heating" multiphysics modules, yielding the final temperature and electric potential distributions. Through the thermoelectric coupling module, Equations (13) and (15) take the following form as,

$$q = -\kappa \nabla T + P\boldsymbol{J} \tag{17}$$

$$\boldsymbol{J} = \sigma_d \boldsymbol{E_d} + \boldsymbol{J_e} \tag{18}$$

Here, $P = ST$ represents the Peltier coefficient, which is the product of the Seebeck coefficient $S$ and temperature $T$, while $\boldsymbol{J_e} = -\sigma_d S \nabla T$ represents the hot carrier current density. Based on the Seebeck effect, the electric potential was calculated as Equations (3) and (4), where Seebeck coeffficient $S$ were 41.8 μV/K for graphene rectangle PPAC array and 44.3 μV/K for unpatterned graphene at $E_f$ = -0.3 eV. PPAC nanodetectors with varying channel lengths connected by two metal electrodes were modeled in COMSOL to calculate the photovoltage. One electrode was grounded, and the electric potential on the other electrode represented the photovoltage, $V_{ph} = V(L_c)$, where $L_c$ is the channel length.

**Acknowledgements**

The authors acknowledge support from the National Key Basic Research Program of China (grant nos. 2024YFA1208500, 2024YFA1208501 and 2025YFA1213200), the National Natural Science Foundation of China (grant no. 92463308), Guangdong Basic and Applied Basic Research Foundation (grant no.2023A1515011876) and China Postdoctoral Science Foundation (grant no. 2025M780806).